# Commensurate - Incommensurate vortex phase in a nanopatterned superconductor

**Gorky Shaw**[1, †]**, S S Banerjee**[1,*]**, T Tamegai**[2] **and Hermann Suderow**[3]

[1] Department of Physics, Indian Institute of Technology, Kanpur-208016, India
[2] Department of Applied Physics, The University of Tokyo, Hongo, Bunkyo-ku, Tokyo 113-8656, Japan
[3] Laboratorio de Bajas Temperaturas, Unidad Asociada UAM, CSIC, Instituto N. Cabrera, Condensed Matter Physics Center, Facultad de Ciencias Universidad Autónoma de Madrid, E-28049 Madrid, Spain

*Email: satyajit@iitk.ac.in

[†] Present address: University of Liege (ULg), Department of Physics, Experimental Physics of Nanostructured Materials, Sart Tilman, B-4000, Belgium.

**Abstract**. Magneto-optical imaging studies on a high-quality $Bi_2Sr_2CaCu_2O_8$ single crystal partially patterned with a triangular array of holes reveal enhanced flux shielding in the patterned region of the sample. By mapping local magnetic field and shielding current density distributions at different applied magnetic fields and temperatures we determine the regime where pinning from the patterned holes dominates over the intrinsic pinning in the sample. In this regime, the flux density near the center of the patterned region is observed to increase when the applied field is varied from below the matching field to just above it, while significant magnetic field gradients are sustained in the patterned region. Our measurements indicate heterogeneous pinning properties of the vortex population, exhibiting signatures of both weak and strong pinning, in the nanopatterned region of the superconductor.

## 1. Introduction

In pinning-free superconductors vortices are arranged in a well-ordered triangular Abrikosov vortex lattice configuration with a lattice period (inter-vortex separation) of $a_0 \sim \sqrt{\phi_0/B}$, where $\phi_0$ is the magnetic flux quantum and $B$ is the magnetic field. The competition and interplay between vortex - vortex interactions and vortex - pin interactions leads to metastable glassy vortex phases, like the well-known Bragg glass and vortex glass [1,2,3] phases. In a superconductor dissipation arises due to normal cored vortices drifting under the influence of a Lorentz force acting on them whenever current is sent through the superconductor. From applications point of view enhancing pinning strength is important as it helps to strongly localize vortices. Strong localization of vortices enhances the threshold current value of a superconductor (viz., the critical current density) beyond which dissipation appears in the superconductor due to delocalized vortices. Recent studies on superconductors with array of holes with diameter of the size of the superconducting coherence length suggest a different interesting mechanism of vortex localization. Popularly studied artificial pinning centers are antidots, which are cylindrical holes open at both ends patterned in a superconducting film. In the antidot array two different mechanisms of vortex pinning have been identified. One type is the pinning of interstitial vortices present in between the antidots when each antidot in the array is saturated with vortices. The strong localization of these interstitial vortices [4] is brought about by the nucleated interstitial vortex itself inducing a modulation of the superconducting order parameter within the interstitial mesoscopic superconducting patch. Recent studies on superconductors patterned with an array of blind holes (cylindrical holes open only on one end) have exhibited properties associated with significant enhancement in pinning [5,6,7,8]. The presence of periodicity in the artificially generated periodic pinning array induces another interesting effect. Superconductors with artificially generated periodic array of antidots show periodic enhancements in the pinning force experienced by the vortices as a function of applied magnetic field ($H$) [9,10,11,12,13,14,15,16,17]. The periodic enhancements are related to the inter-vortex separation ($a_0$) becoming commensurate with the period ($d$) of the lattice of pins at integral multiples of the first matching field ($B_\phi$), where $a_0 \sim d$, $B_\phi(a_0 = d) \sim \frac{\phi_0}{d^2}$. As a function of the $H$ a variety of different commensurate and incommensurate phases appear.

In superconductors with a natural random distribution of pinning a well-known critical state exists viz., the Bean critical state [18]. The Bean critical state is an example of organization setting in a system out of equilibrium. An example of self-organized critical state is well known for sand-piles, wherein there is a critical slope of the sand-pile. The addition of sand to enhance the slope beyond this critical value leads to destabilizing the pile and the generation of an avalanche until the slope returns back to its critical value. The vortex configuration attained by increasing the field is not a system in equilibrium. The presence of pinning centers which localize vortices keeps the vortex configuration from achieving its ground state configuration. In superconductors with random distribution of pinning centers, when the applied field is increased from zero to a finite field value, the penetrating vortices organize themselves into a configuration with a gradient in density. The vortex density profile from the sample edge to the center has a critical slope akin to the sand-pile. The vortex density decreases from the edge to the center of the superconductor. The vortex density gradient from all the sample edges results in a nonzero curl of the local field $\vec{B} \propto$ current density ($J$). According to the Bean critical state model the value of this shielding current density $J$ is equal to the critical current density ($J_c$) of the superconductor (which is a measure of the pinning strength in the superconductor). The diamagnetic shielding currents lead to a uniform gradient in the magnetic field distribution across the superconductor. Thus external magnetic field sets up diamagnetic shielding currents which determine the magnetization response of the superconductor. The diamagnetic shielding currents set up in the superconductor are in turn related to field distribution in the superconductor which we investigate in this paper.

In superconductors with periodic arrangement of pins, the Bean like state is expected to be modified as a result of a balance [19,20] between two opposing tendencies. One tendency is associated with forming an ordered arrangement of vortices commensurate with the periodic lattice of pins and thereby generating a uniform (gradientless) field distribution. The competing tendency favours a uniform constant gradient in the field distribution or a constant bulk critical current density in the superconductor. The competition between these tendencies leads to domains of vortices which are ordered around the periodic pinning sites. Unlike the antidots, blind holes are relatively weaker pins, where the vortex - blind hole interaction does not overwhelm the inter-vortex interaction inside the superconductor. In superconductors patterned with relatively weak pinning blind holes the inter-vortex and vortex – weak pin interactions are comparable in magnitudes. Thus such a system is potentially favourable to study a variety of different vortex configurations arising out of competition and interplay between the sample's intrinsic pinning and the pinning from patterned blind holes.

In this paper we investigate a triangular array of blind holes patterned on a superconducting single crystal instead of thin films to ensure intrinsic pinning in the superconductor doesn't dominate the vortex pin interaction. Furthermore we restrict our study to low $B < B_\phi$, viz., the dilute vortex density case to keep the inter-vortex interaction low. For this situation at first glance it may be expected that due to comparable inter-vortex and vortex - blind hole pin interaction, vortices will be uniformly distributed across the blind hole array with a uniform filling fraction of $\sim B/B_\phi$. In such a vortex configuration on an average only $B/B_\phi$ of the total number of blind holes will be occupied with single vortices. In this work using Magneto-Optical Imaging (MOI) technique we image the nature of field distribution in the blind hole array patterned on a $Bi_2Sr_2CaCu_2O_8$ (BSCCO) single crystal. MO imaging of the magnetic flux distribution over the patterned region reveals enhanced flux shielding in the patterned region. We have mapped the local magnetic field and shielding current density distributions at different applied magnetic fields and temperatures. Using these we determine the regime where blind hole pinning dominates over the intrinsic pinning. In this regime, the effective local magnetic flux penetration field is found to be higher in the patterned region. It is observed that in this scenario, when the applied field is varied from below the matching field to just above it, the flux density near the center of the patterned region increases while significant magnetic field gradients are sustained in the patterned region. Our measurements indicate heterogeneous pinning properties of the vortex population in the nanopatterned region of the superconductor, with features suggestive of both weak pinning and strong pinning inside the patterned region. We suggest a simple model to explain features in our data, as well as compare our results with predictions from more rigorous recent theories, which are partially related to our work.

## 2. Experimental details

*2.1. Sample preparation*
For our experiments we use a high-quality single crystal of $Bi_2Sr_2CaCu_2O_8$ (BSCCO) [5,21] of dimensions ($0.8 \times 0.5 \times 0.03$ mm$^3$) and $T_c$ = 90 K (figure 1(a)). The sample surface ('*ab*' plane of the single crystal) was milled using Focused Ion Beam (FIB) machine (dual beam FEI make Nova 600 NanoLab) with a focused Ga ion beam (diameter $\sim$ 7 nm) to create a triangular array of circular blind holes covering an area of $\sim$ 40 µm $\times$ 40 µm. Figure 1(b) shows a magnified Scanning Electron Microscopic (SEM) image of the blind hole array. Mean diameter of the holes is 170 nm, mean center-to-center spacing between the holes (*d*) is 350 nm (BSCCO penetration depth $\lambda_{ab} \sim$ 200 nm for '*ab*' crystal orientation) and depth of the holes is $\sim$ 500 nm (thickness of single crystal = 30 µm). For this blind hole array the matching field is $B_\phi \approx 195$ G.

*2.2. Magneto-Optical Imaging*
The sample was studied using a high-sensitivity magneto-optical imaging (MOI) technique [5,21,22]. In the MOI setup a sensitive CCD camera captures the Faraday light intensity distribution, *I*(*x, y*), across a superconductor placed in a constant applied magnetic field $H_z$ (applied parallel to the *c*-axis of the single crystal). Linearly polarized light reflected off a Faraday active garnet film with a high Verdet constant (*V*) placed in close contact with the sample surface helps produce a distribution in the Faraday rotated light intensity reflected off different regions on the sample surface where there is a distribution in the local magnetic field $B_z(x, y)$. Recall the Faraday rotation angle $\theta(x, y) \propto VB_z(x, y)$.

The reflected Faraday rotated light intensity contains information of $B_z(x, y)$ across the sample surface (Faraday rotated light intensity $I(x, y) \propto (B_z(x, y))^2$, $\hat{z}$ is an orientation perpendicular to the crystal surface which is the *ab*-plane of the single crystal). In our measurements we also perform differential MOI (DMOI) [23,24] wherein, differential Faraday rotated light intensity, $\delta I(x, y)$, is captured in response to step increase in the $H$, $\delta H = 1$ Oe, (viz., $\delta I(x, y) = <I(x, y)$ at $H + \delta H> - <I(x, y)$ at $H>$, where <..> represents averaging over $n$ (= 20) images at each $H$). By measuring $\delta I(x, y)$ in the DMOI images produced in response to a calibrated change in the $H$ far away from the sample being investigated, we determine the local magnetic field changes $\delta B_z(x, y)$ across the superconductor produced in response to $\delta H = 1$ Oe. Here we emphasize that due to irreversible magnetization response within the nanopatterned region we do not repeatedly modulate $\delta H$ about $H$, , which is unlike conventional DMOI technique [23,24].

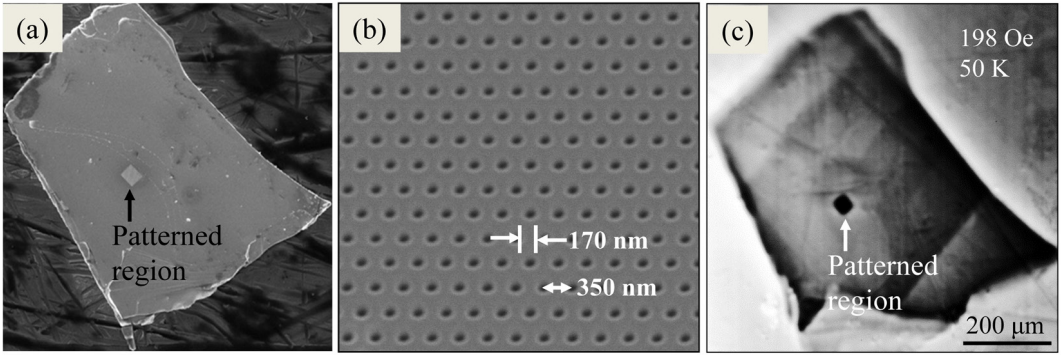

**Figure 1.** (a) Scanning electron microscope (SEM) image of the sample. The patterned region is the bright rectangular area (~ 40 μm × 40 μm) indicated by the arrow. (b) Magnified portion of the patterned region, showing the triangular array of blind holes of diameter 170 nm and with center to center hole spacing of 350 nm. Depth of the holes is ~ 500 nm. The triangular array of holes correspond to a matching field $B_\phi = 195$ G. (c) Conventional MO image of the zero field cooled sample (ZFC) at $H = 198$ Oe and $T = 50$ K. The patterned region is the dark rectangular area indicated by the arrow.

Figure 1(c) shows a conventional MO image showing the full sample to identify different features in such an image. This image was obtained at $H = 198$ Oe and $T = 50$ K after zero field cooling (ZFC) the sample. The contrast variations in the image represent changes in $B_z(x, y)$ across the sample. In this MO image the nanopatterned region is readily identified by the square shaped region with dark contrast compared to the surrounding brighter contrast in the unpatterned regions of the sample. The darker contrast inside the patterned region represents a lower average $<B_z(x, y)>$ inside the patterned region compared to the brighter contrast in the unpatterned regions.

## 3. Results and discussion

### 3.1. Local isothermal magnetization response

We study the local isothermal magnetization ($M(H)$) response in different regions of the sample using conventional MOI (viz., we use conventional MOI to obtain the $B_z(x, y)$ distribution in the sample without using $\delta H$ modulation). The $M(H)$ behaviour is deduced using $4\pi M(H) = (\int_A [B_z(x, y) - H] dxdy)/A$, where $A$ is the area over which averaging is performed and the local field $B_z(x, y)$ is determined from conventional MO images. Typically, $A = 25$ μm$^2$. Figure 2 shows $M(H)$ curves for the nanopatterned region (red, squares) and the whole sample (bulk, using $A$ = entire sample area (blue, circles)), at different $T$. Clearly, the local $M(H)$ behaviour in the patterned region is significantly different from the bulk response at lower $T$, and at higher $T$ the two magnetization responses merge with each other.

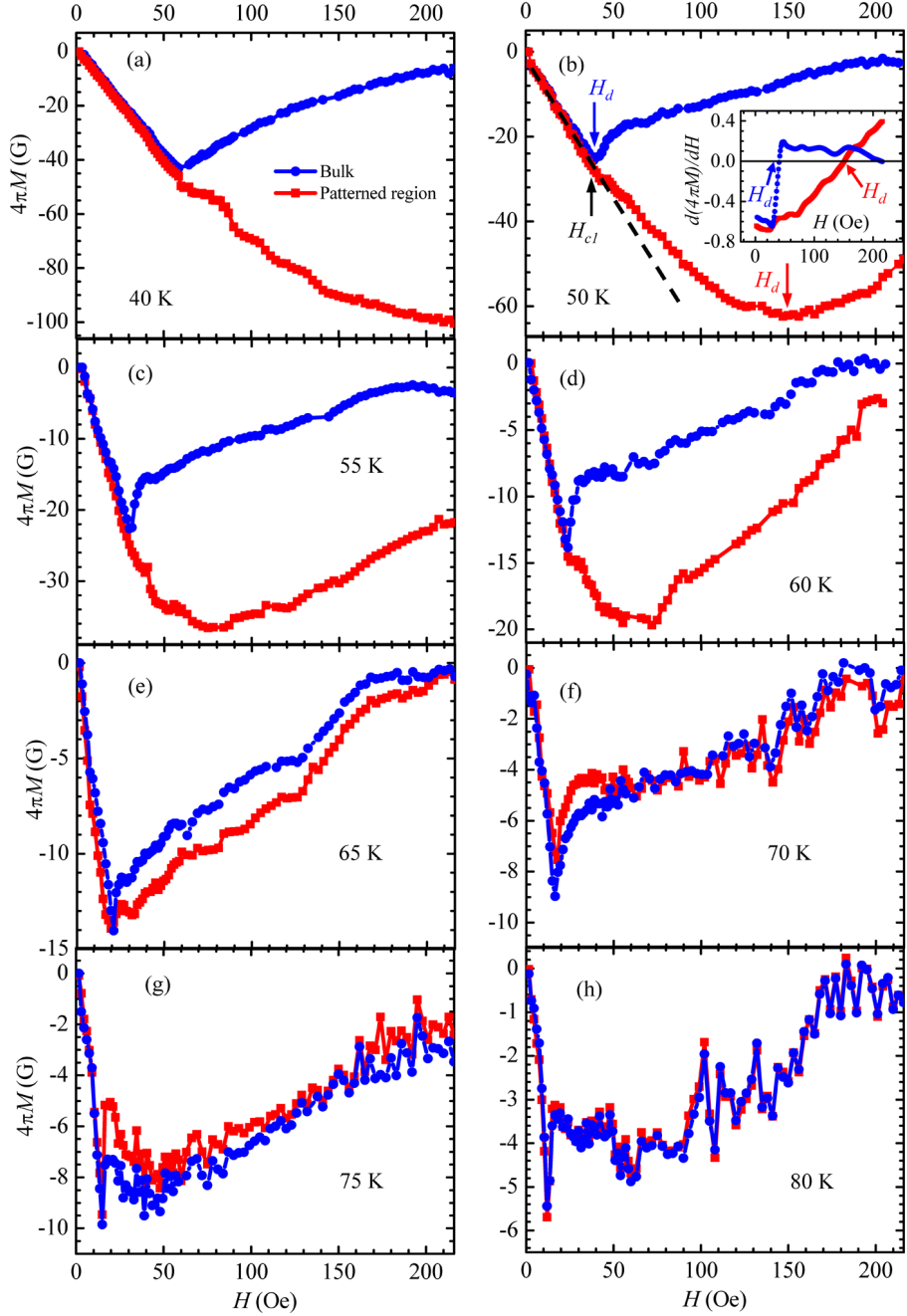

**Figure 2.** (a)-(h) $M(H)$ curves obtained from MO images at different $T$ as indicated in each panel. The blue curves (circles) represent $M(H)$ response for the entire sample area. The red curves (squares) represent local $M(H)$ response in the patterned region. Inset of (b) shows $dM/dH$ vs. $H$ plots at 50 K obtained from the $M(H)$ curves shown in the main panel. The mean penetration fields for the patterned region and the bulk at 50 K are indicated in (b) by the dashed and dotted arrows, respectively.

We further analyse the $M(H)$ responses at 50 K. The dashed line in figure 2(b) identifies the location in the vicinity of which the slope of the bulk $M(H)$ curve changes from linear behaviour. The field ($H$) up to which the linear $M(H)$ behaviour is maintained identifies the lower critical field ($H_{c1}$) of the superconductor. At $H < H_{c1}$ the local field $B_z = 0$ G inside the superconductor (Meissner state), due to which $4\pi M = -H$ (up to demagnetization factor), viz., a linear behaviour in $M(H)$.

At $H > H_{c1}$ the $M(H)$ begins to deviate from the linear behaviour and the negative slope ($dM/dH$) which was at its maximum value below $H_{c1}$ begins to decrease. At a higher $H$ the slope changes from negative to positive value as shown in $dM/dH$ vs. $H$ plots in inset of figure 2(b). At $H > H_{c1}$ the vortices begin penetrating into the superconductor, however they do not achieve a uniform distribution. There are gradients in the local vortex distribution and the center of the sample (or patterned region in our case) is free of vortices up to a field $H_d$. The regime where the $dM/dH$ changes curvature helps us identify the location of the mean penetration field $H_d \sim 40$ Oe at 50 K where vortices have penetrated into the bulk of the unpatterned regions of the sample. By noting the deviation from linearity in $M(H)$ the bulk $H_{c1}$ is estimated to be $\sim 40$ Oe (at 50 K), and it is found to be almost the same for the patterned region also. The local $M(H)$ curve of the patterned region is also found to be linear up to the local $H_{c1} \sim 40$ Oe (indicated by the black arrow in figure 2(b)). However the local magnetization curve for the patterned region shows enhanced diamagnetic response due to enhanced pinning compared to the unpatterned regions of the sample. This in turn indicates that the patterned region is associated with enhanced screening currents [18], which are stronger than those present in the unpatterned regions. However, even beyond 40 Oe, the deviation is from linearity is gradual and the $M(H)$ response is strongly diamagnetic. The gradual deviation from linearity indicates that the flux has not penetrated fully up to the center of the patterned region. Above $H_d$, the diamagnetic magnetization response monotonically decreases with increasing $H$ as the flux has reached the center of the superconducting region under consideration. The dotted arrow in figure 2(b) identifies the penetration field $H_d \sim 150$ Oe (determined from the $dM/dH$ vs. $H$ plot in inset of figure 2(b)) at 50 K for the 25 $\mu m^2$ area inside the patterned region as compared to a value $\sim 40$ Oe in the unpatterned regions of the sample. As $H_d \propto J_c$ [18], the ratio of $H_d$ at 50 K shows $J_c$ of the patterned region is at least three to four times larger than that in the unpatterned regions.

Note from figure 2 that with increasing $T$ the mean penetration field $H_d$ for the patterned region decreases from 150 Oe at 50 K to 72 Oe at 60 K. $H_d$ for both curves are identical at higher $T$ as the local and bulk magnetization responses merge with each other. From these observations, we can assert that the effective pinning strength of the patterned region is $\sim 70$ K. For $T < 70$ K, pinning from the patterned blind holes dominates over the intrinsic pinning of the sample (as observed in the enhanced flux shielding inside the patterned region in this regime). We note that at 60 K, local $M(H)$ response of the patterned region indicates higher flux shielding inside the patterned region compared to the bulk of the sample and approaches the bulk $M(H)$ curve at higher $H$. Hence, for further analysis, we choose 60 K as the regime of interest. At 60 K the pinning properties of the patterned blind holes are prominent, while not being overwhelming, and the intrinsic pinning in the unpatterned regions of the sample are weakened. We further investigate local shielding response inside the patterned using DMOI.

*3.2. Measurement of the local shielding response using DMOI*
We now investigate the nature of the field gradients inside the patterned region and how susceptible the gradients are to field changes. Shown in figures 3(c) and 3(d) are DMOI images of the patterned BSCCO sample. To aid in understanding the DMOI images in figures 3(c) and 3(d), we show two schematics in figures 3(a) and 3(b). The images at the bottom of the figures 3(a) and 3(b) show schematic DMOI images showing two regions with two different gray shades which correspond to a larger change (light gray) and a smaller change (dark gray) in $B_z$ in response to field change $\delta H = 1$ Oe, respectively. The main panels in figures 3(a) and 3(b) show two possible Bean like critical $B_z(r)$ profiles across the white dashed line drawn in the insets of figures 3(a) and 3(b). In these figures, the black and red curves represent the $B_z$ profile at $H = 10$ and $H + \delta H = 11$ (arb. units), respectively. In figure 3(a) main panel the difference between the $B_z(r)$ profiles across the region marked unpatterned is larger than that in the region marked patterned. Therefore in a differential image where the contrast in the image is proportional to the difference in $B_z$ at $H$ and $H + \delta H$, one obtains a comparatively darker contrast within the patterned region (cf. figure 2(a) inset). The situation is opposite in figure 3(b).

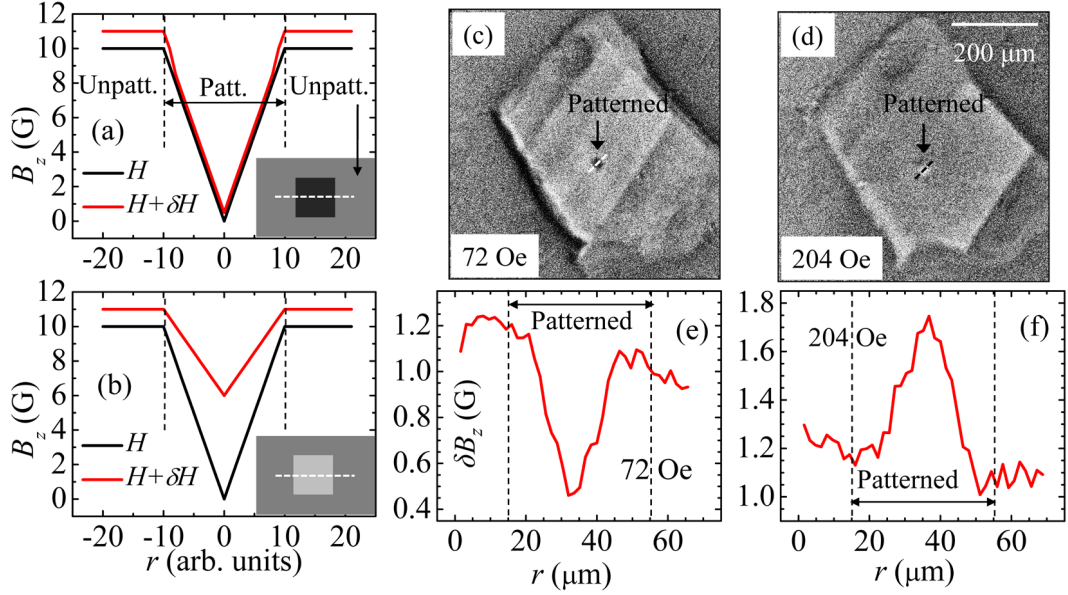

**Figure 3.** (a) and (b) main panels are schematics representing possible behaviour of $B_z(r)$ profiles across the patterned region at applied fields $H$ and $H + \delta H$ (black and red curves, respectively) across a region corresponding to the white dashed line drawn in the respective insets. The insets of (a) and (b) show the schematic differential magneto-optical image (DMOI) where the contrast difference in the image is obtained by taking the difference between the $B_z(r)$ profiles at $H$ and $H + \delta H$ in the (a) and (b) main panels, respectively. The different shades of gray correspond to a larger difference (light gray) and a smaller difference (dark gray) in $B_z(r)$ in response to $\delta H = 1$ Oe, respectively. (c) and (d) are actual DMOI images measured at 60 K with $H = 72$ and 204 Oe, respectively. (e) and (f) are the $\delta B_z(r)$ behaviour measured across the patterned region at 60 K at $H = 72$ and 204 Oe, corresponding to (c) and (d), respectively. The patterned region is identified by the dashed vertical lines in (e) and (f).

Figures 3(c) and 3(d) shows actual experimental DMOI images of the patterned region of the sample captured at 60 K and $H = 72$ and 204 Oe, respectively. Figures 3(e) and 3(f) show the $\delta B_z(r)$ measured along the dashed lines in figures 3(c) and 3(d), respectively. We can see that by comparing figures 3(e) and 3(a), that the $B_z(r)$ profile (viz., the vortex distribution) inside the patterned region doesn't get significantly altered due to $\delta H$ of 1 Oe when compared to the unpatterned regions of the sample. From figures 3(e) and 3(f) we observe that outside the patterned region the $\delta B_z(r) \approx \delta H = 1$ Oe, indicating nearly reversible response in these unpatterned regions of the sample. In comparison to figure 2(c), the situation changes quite dramatically at higher fields, e.g. at 204 Oe (cf. figures 3(d) and 3(f)). A comparison of figures 3(f) and 3(d) at 204 Oe shows significantly larger changes present in vortex density distribution inside the patterned region, leading to a bright contrast inside the patterned region in response to the $\delta H$ modulation. From the conventional Bean model [18], as compared to field distribution at low $H$, at higher $H$ the average $B_z$ inside the superconductor is higher. Also, the gradient in the field distribution decreases w.r.t. gradient at lower $H$ since the vortex distribution approaches closer to a uniform distribution with increasing inter-vortex interaction. Hence the brightening observed in figure 3(d) implies that at higher fields the vortex state inside the patterned region is approaching the expected behaviour in a weak pinning superconductor. Note that in the unpatterned region of the superconductor as the sample is reversible with $B_z \sim H$ hence the maximum $\delta B_z$ change expected in the local field will be of the order of $\delta H$. Note the appearance of bright contrast inside the patterned region in DMOI images begins only above 132 Oe. It appears that at low fields when the inter-vortex interaction is low and comparable with the vortex - blind hole pinning the dilute vortex configuration is quite stable.

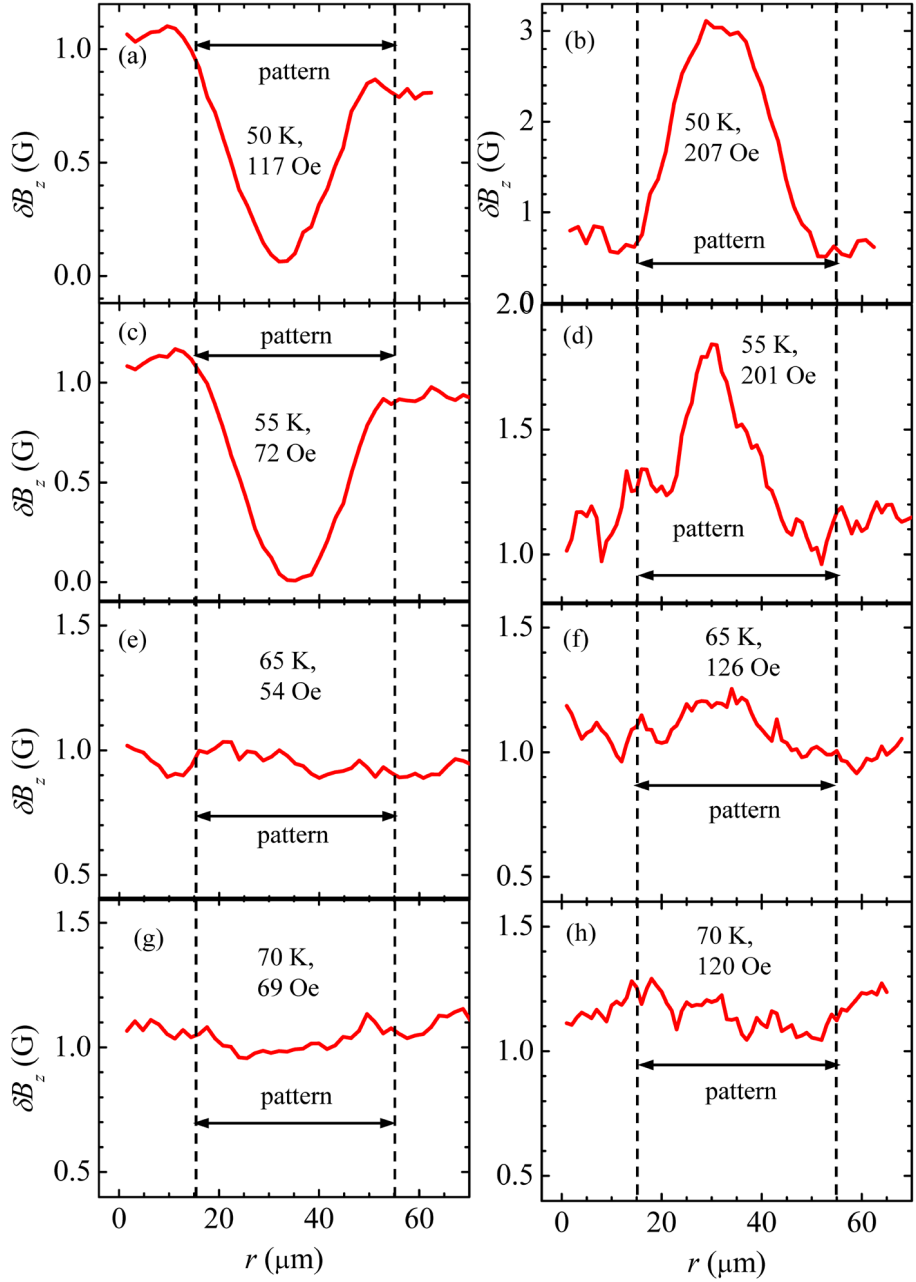

**Figure 4.** $\delta B_z(r)$ behaviour across the patterned region at various $T$ above and below 60 K. (a)-(b) 55 K, (c)-(d) 60 K, (e)-(f) 65 K and (g)-(h) 70 K, at different $H$ as indicated in each panel. The patterned region is identified by the dashed vertical lines in (a)-(h).

Figure 4 shows the $\delta B_z(r)$ measured across the patterned region in similar fashion as in figure 3 at different $T$ above and below 60 K. As expected from the observations in figure 2, the $\delta B_z(r)$ profiles at 50 K and 55 K (< 60 K) (cf. figures 4(a)-(d)) follow similar trends of lower $\delta B_z$ inside the patterned region transforming to higher $\delta B_z$ as $H$ is increased, as observed at 60 K (cf. figure 3). And at 65 K and 70 K (> 60 K) (cf. figures 4(e)-(h)), the $\delta B_z$ inside the patterned region is seen to be almost indistinguishable from that outside it. By comparing figures 2, 3 and 4 we see that at each $T$ near the local penetration field $H_d$ for the patterned region the DMOI response ($\delta B_z$) changes from small to larger values inside the patterned region in comparison changes in rest of the sample, i.e., $\delta B_z$ inside the patterned region changes from ~ 0 G to value greater than $\delta H$ (the field modulation = 1 Oe).

*3.3. Gradients in vortex density in the vicinity of the patterned region*

To further analyse the magnetization response, we obtain the variation of local field $B_z(x, y)$ profile with $H$ across the nanopatterned region determined from conventional MOI at different $T$. Figures 5(a)-5(d) show maps of the $B_z(x, y)$ (as deduced from MO measurements) in and around the patterned region shown with a resolution of 0.1 G at $T = 60$ K and $H = 51, 96, 144$, and 216 Oe, respectively. The patterned region is indicated by the dashed square in figure 5(d) for clarity. The scale bar beside each image represents the variation in $B_z(x, y)$ in the corresponding image. Note that the intensities (contrast) in a conventional MO image represent variations in the distribution of $B_z$, unlike a DMOI image in which the intensities represent variations in $\delta B_z$ produced in response to a $H$ modulation (cf. discussion on MOI technique in section 2.2). We note that in each image, the lowest $B_z$ value (darkest) is observed near the center of the patterned region, and highest $B_z$ value (brightest) is observed outside its periphery. Figure 5(e) shows $B_z(r)$ profiles across the patterned region (along the dashed line shown in figure 5(a)) at 60 K, at different $H$ as indicated along the profiles. The vertical dashed lines in figure 5(e) indicate the boundaries of the patterned region. It is observed that as $H$ is increased; significant gradients (slope) in the $B_z$ profile start developing across the patterned region.

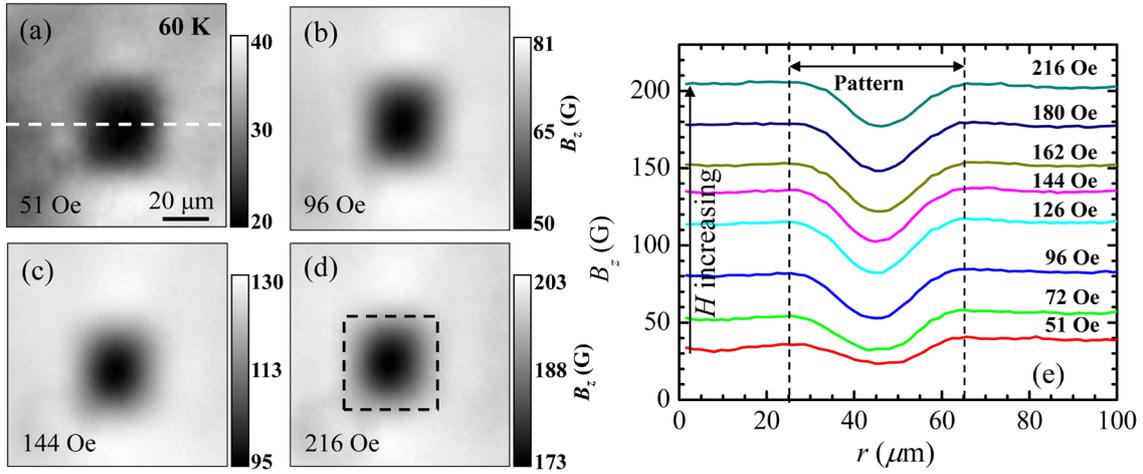

**Figure 5.** (a)-(d) Contour maps of the $B_z(x, y)$ distribution in a ~ 80 μm × 80 μm region around the patterned region at 60 K and $H = 51, 96, 144$ and 216 Oe, respectively. The patterned region is indicated by the dashed square in (d) for clarity. (e) $B_z(r)$ profiles across the patterned region along the dashed line shown in (a), at 60 K at different $H$ as indicated on the figure. The patterned region is identified by the dashed vertical lines in (e).

The presence of significant large gradients around the patterned region suggests that any change in $B_z$ inside the patterned region in response to a change in $H$ is shielded by the patterned region. This is consistent with what has been discussed previously in context of figures 2 and 3. However, from the scale bars, it is noted that at $H <$ and close to $B_\phi$, the local $B_z$ inside the patterned region is non-zero and it increases with $H$. This indicates that the patterned region does not completely shield magnetic field changes even at lower $H$. Note that the $B_z(r)$ data figure 5(e) corresponds to the particular line along which the profile has been obtained. However the same might not be the case along all the directions of the patterned area, as clearly from the images in figures 5(a)-5(d) $B_z$ distribution across the patterned region is not uniform (cf. the contrast variation along the boundary of the patterned region indicated by the dashed square in figure 5(d)). At 60 K, the gradients in the $B_z$ profiles remain almost unchanged for higher $H$ ($> B_\phi$) and are sustained up to the maximum $H$ applied in our experiments. From the above observations, we conclude that the nanopatterned region is characterized by an unusual heterogeneous pinning response. On one hand it exhibits characteristics of strong pinning, with significant gradients in $B_z$ being sustained near the periphery (cf. figure 5), while local $B_z$ increases with relative ease with increase in $H$ near the center (cf. figures 3 and 5), which is characteristic of weak pinning.

*3.4. Mapping the distribution of shielding currents*

By numerically inverting the field distribution [25,26] measured across the patterned region using MOI we determine the absolute value of the screening current distribution $|j(r)|$ inside the patterned region at 60 K. Figures 6(a)-6(d) show maps of the $|j(r)|$ distribution in and around the patterned region.

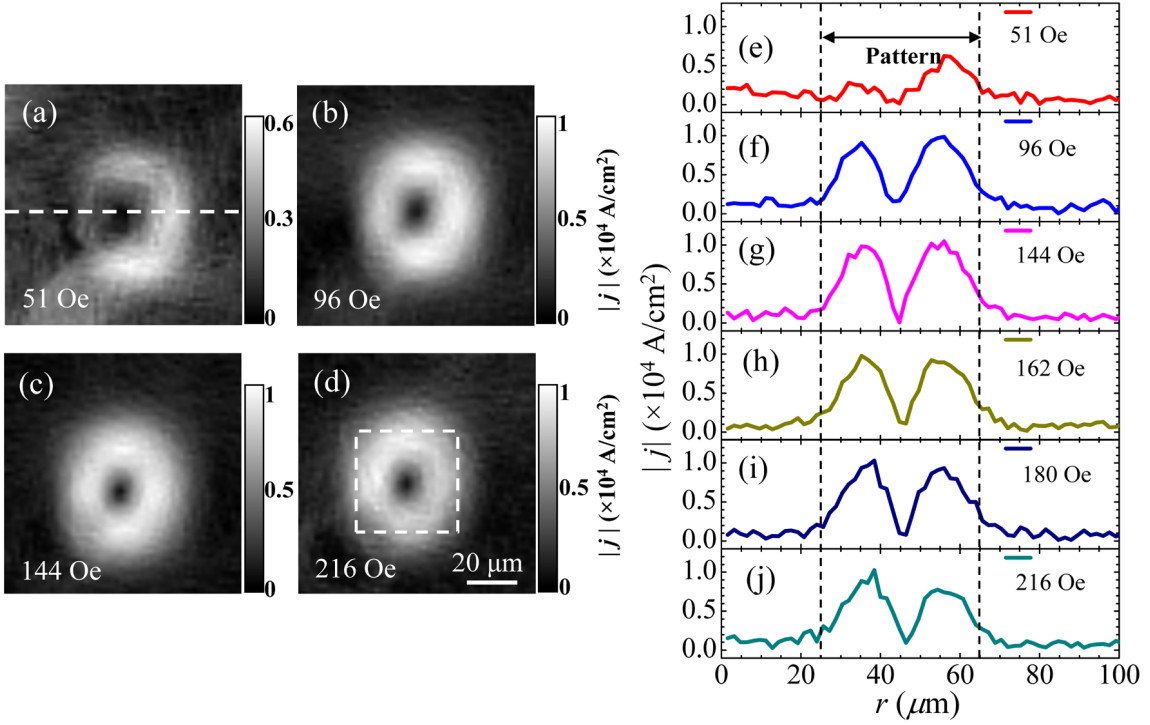

**Figure 6.** (a)-(d) Contour maps of the $|j(r)|$ distribution, obtained from the $B_z(x, y)$ distribution, in a ~ 80 μm × 80 μm region around the patterned region at 60 K and $H$ = 51, 96, 144 and 216 Oe, respectively. The patterned region is indicated by the dashed square in (d) for clarity. (e)-(j) $|j(r)|$ profiles across the patterned region along the dashed line shown in (a), at 60 K at different $H$ as indicated on the figure. The patterned region is identified by the dashed vertical lines in (e)-(j).

The patterned region is indicated by the dashed square in figure 6(d) for clarity. The graphs in figures 6(e)-(j) show the $|j(r)|$ profiles along the dashed line (cf. figure 6(a)) at different $H$. The scale bar along each image in (a) to (d) indicates the range of $|j|$ in the corresponding image (in units of $10^4$ A/cm$^2$). At $H$ = 51 Oe, (cf. figure 6(a)) large $|j(r)|$ around the periphery of the patterned region (indicated by the bright contrast in these regions) is observed. This corresponds to these regions possessing relatively high gradients in $B_z(r)$. Far outside the patterned region $|j(r)|$ is almost zero (indicated by the dark contrast in these regions), indicating that $B_z(r)$ distribution in the unpatterned regions of the sample is almost constant. As $H$ is increased, $|j(r)|$ at the boundary of the patterned region increases, as indicated by the appearance of bright contrast over larger regions, and also higher value corresponding to the bright contrast in figures 6(b)-6(d). Also note that the $|j(r)|$ distribution is not uniform across the patterned region. At 51 Oe, larger $|j(r)|$ is observed along the lower right edge of the patterned region, whereas at higher $H$, larger $|j(r)|$is observed along the upper right edge. These features are also seen in the $|j(r)|$ profiles in figures 6(e)-(j). At 51 Oe (cf. figure 6(e)), $|j(r)|$ near the right edge of the patterned region is higher than the left edge. Subsequently at higher $H$ (cf. figures 6(f)-(j)), the overall $|j(r)|$ values near the periphery of the patterned region are enhanced and the large $|j(r)|$ values are sustained at higher $H$. These observations are consistent with those in figure 5 above as gradients in $B_z(r)$ are proportional to $j(r)$ [18] (Note that $|j(r)|$ represents the absolute value only, and not the sign of *j*).

*3.5. Heterogeneous pinning in the patterned region*

As discussed above, we observed in figures 3-6 that with increasing *H* there is an increase in $B_z$ near the center of the patterned region while large gradients near the periphery of the patterned region are sustained. $B_z(r)$ and $|j(r)|$ distributions inside the nanopatterned region are also seen to vary non-uniformly across the patterned region (cf. figures 5 and 6) at low *H*. These results suggest that the vortex population inside the nanopatterned region of the sample exhibits signatures associated either with weak or strong pinning viz., a heterogeneous pinning response at low *H* below $B_\phi$. We believe that the heterogeneous pinning response indicates an unusual distribution of vortices inside the patterned region, contrary to the expectation that at $H < B_\phi$ vortices are mostly pinned on blind holes.

In the absence of a general model which allows a direct comparison with the peculiar nature of field gradients and current distribution determined in our nanopatterned pinning array study (specifically in the specific *H-T* regime we investigate), we present a speculative one dimensional microscopic model. The proposed microscopic model is a simplified one dimensional model which is similar to a 1-d model proposed to understand general features commensurate incommensurate states in condensed matter systems [27]. We adapt this model [27] to describe a one dimensional distribution of vortices over a one dimensional periodic array of pins (cf. figure 7(a)). In its present form the model helps make qualitative connections with the data. The model presently lacks rigor for direct quantitative comparisons with maps of field distributions over a pinning array in effectively two dimensions. We hope future work will help evolve the model allowing for more quantitative comparisons with actual data. Infact a recent theoretical study [20] has considered the peculiarities of the vortex distribution in mesoscopic superconductors with correlated pins. While this study [20] doesn't directly allow for a direct quantitative comparison with the gradients we measure however there are interesting inferences one can draw from this work which we will discuss subsequently.

Returning back to the model we propose: In this periodic configuration of pins the vortices are expected to separate into a commensurate - incommensurate (CI) vortex state as shown in figure 7(a). The vortices are represented as red circles and the periodic pinning potential created by periodic array of pins (or patterned holes as in our experiment) as a solid wavy line. If the periodic pinning potential has a periodicity *b*, then the vortex location in the CI phase at the $n^{th}$ pinning site is described in terms of the inter-pin distance *b* by a phase term [27] $\varphi_n$, $x_n = nb + \dfrac{2\pi}{b}\varphi_n$. In a continuum limit by considering *x* as a continuous variable one can write a differential equation for $\varphi(x)$ [27]. The one dimensional differential equation has a Solitonic solution for $\varphi(x)$. We describe the generation of an incommensurate vortex state separating the commensurate vortex phase as a domain wall (DW) in terms of the Solitonic solution [27] with the phase $\varphi(x)$ behaving as $\varphi(x) \propto \tan^{-1}[\exp(x)]$. The phase, $\varphi(x)$ changes from 0 to $2\pi$ from left to right across the domain wall as shown in figure 7(b) as per $\varphi(x) \propto \tan^{-1}[\exp(x)]$ (note, $\varphi(-x)$ is an equally possible solution, with the $\varphi(x)$ changing from 0 to $2\pi$ from right to left). Thus the incommensurate region in the vortex state is identified with the nucleation of a Domain wall in the system as shown in figures 7(a) and 7(b).

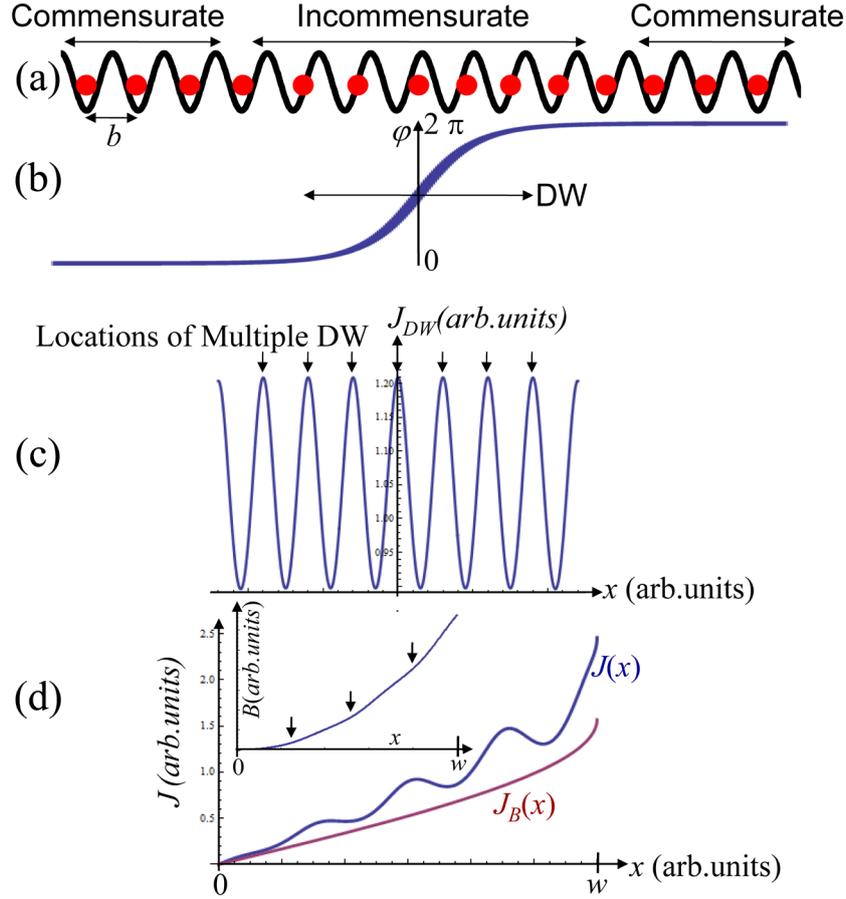

**Figure 7.** (a) A schematic representation of commensurate – incommensurate (CI) phase of vortices (represented as red circles) on a periodic pinning force background (black solid lines). (b) The behaviour of the phase $\varphi(x)$ describing the domain wall (DW) a CI phase, (c) Current distribution profile associated with multiple domains walls ($J_{DW}$) in the one dimensional model. (d) Current density profile $J(x)=J_{DW}(x)J_B(x)$. For comparison we show the Bean current density profile $J_B(x)$. In the inset we show the $B(x) = \int_0^x J(x)dx$ behaviour calculated numerically from the $J(x)$. The location of breaks in slope of $B(x)$ is indicated by arrows. In the graphs shown, $w$ represents the width of the patterned region from its center.

From figure 7(a) it is clear that across the DW, viz., in the incommensurate region, there will be a change in the density of vortices and hence a magnetic field gradient will develop in this region compared to the region where there is a commensurate phase (the local field distribution here will be is uniform). We propose the behaviour of the vortex density, as one moves from the commensurate to the incommensurate region in figure 7(a), can be modelled as the vortex number density behaving as $\rho[x] = \delta(x-nb) + c_{incom}(1-Cos[\phi(x)])$ where the $\delta(x-nb)$ represents a delta function for the number density at each pin site in the commensurate region and $c_{incom}$ is a positive constant and the term gives the extent of change in number density in the incommensurate DW region. In the incommensurate region the factor $(1-Cos[\varphi(x)])$ is zero near the edges of the DW or the incommensurate region where $\varphi(x) = 0$ or $2\pi$ and is maximum near the center where $\varphi(x) = \pi$.

We now attempt to estimate the slope of the local vortex density at a given point inside DW: Across the DW the density of vortices changes. The incommensurate DW region is characterised by change in $\varphi$, therefore around a given value of $\varphi(x)$ we estimate the local slope of $\varphi(x)$ and hence the changes in vortex density. For an infinitesimal $\delta\varphi$ increment in $\varphi$ at a point inside DW, corresponds to an incremental change in $(1-Cos[\varphi(x)]) \sim 2\varphi(x)\delta\varphi$. This in turn leads to a change in local vortex density at a point inside the DW, by an amount $\delta\rho[x] \sim c_{incom}\phi(x)\delta\phi$. Note as in general $B[x] = \phi_0\rho[x]$ is valid, therefore from this 1-D model one may determine the Bean like currents associated with local changes in gradients using $dB/dx \propto d\varphi/dx$. From the above discussion one can find that the local current density at a location inside a domain wall ($J_{DW}$) by $J_{DW} \propto \frac{dB}{dx} \propto \frac{d\varphi(x)}{dx} \propto \frac{d}{dx}\{\tan^{-1}[\exp(x)]\}$ (here we have considered the dominant contribution to Bean critical current as coming from the DW region).

Above we have considered only one DW, next we consider multiple domain walls which can be excited in this system. We model the current density associated with equally spaced multiple domain walls as

$$J_{DW}(x) \propto \frac{d}{dx}\{\sum_k \tan^{-1}[\exp(x-k)]\} \qquad (1)$$

where $k$ is an integer index identifying the location of a domain wall. For multiple domain walls in this 1-D model the $J_{DW}$ is shown in figure 7(c), where the different peaks located with arrows are associated with current peaks at domain walls at the location index $k$. If the domain walls are reasonably well separated then the modulation in $J_{DW}$ is much larger as compared to that shown in figure 7(c). Apart from $J_{DW}$, inside the patterned region there is an overall Bean current density profile which is of a well-known form obtained from the modified Bean profile [28]

$$J_B(x) \propto \tan^{-1}\left[\frac{x}{\sqrt{w^2-x^2}}\right] \qquad (2)$$

where $w$ is half the width of the patterned region. Using equations (1) and (2), we get overall current density profile inside the patterned region as

$$J(x) \propto \frac{d}{dx}\left\{\sum_k \tan^{-1}[\exp(x-k)]\right\} \times \tan^{-1}\left[\frac{x}{\sqrt{w^2-x^2}}\right] \qquad (3)$$

as the overall Bean current density profile modulated by the domain wall current profile ($J_{DW}$). The behaviour of $J(x)$ is shown in figure 7(d) along with the behaviour of $J_B(x)$ also shown on the same plot for comparison (note in figure 7(d), $w$ represents the sample width, and we plot the behaviour for only half the sample). It is clear from figure 7(d) that the current profile inside the patterned region gets modulated by the presence of multiple domain walls and the current profile appears uneven and rides above the Bean profile. Note from figure 7(d) that the $J(x)$ appears higher than the conventional Bean like $J_B(x)$. The reason for this is that we have considered very closely spaced domain walls due to which there is a significant overall contribution of the currents from the domain walls in the $J(x)$ expression and therefore the average current density increases. By considering sufficiently well spaced out DW's the above feature does not appear. From the above model, it appears at the location of a domain wall there is a peak in the current density $J(x)$.

Using equation (3), we calculate the behaviour of the magnetic field profile $B(x)$ in the one dimensional system, by numerically integrating, $B(x) \propto \int J(x)dx$. The behaviour of $B(x)$ is shown in figure 7(d) inset. As expected the peaks in the $J(x)$ naturally produce breaks in the slope of the magnetic field distribution $B(x)$ (cf. location of arrows in figure 7(d) inset). From figure 7(d) it can be summarized that while the breaks in slope of $B(x)$ due to DW is less easy to discern, their evidence in $J(x)$ behaviour is seen more strongly. Due to the presence of domain walls inside the patterned region there is a distribution of current densities inside the patterned region and the Bean like profile gets modified. The above is characteristic of heterogeneous pinning properties which can be found due to CI transition in the vortex state in the presence of nanopatterned pinning array.

Based on the above calculations, the heterogeneous pinning inside the patterned region may arise from domain wall boundaries excited in the vortex configuration inside the nanopatterned region of the sample. A detailed recent theoretical study [20] also suggests a nested domain like structure for the vortex configuration within a mesoscopic superconductor with correlated pinning centers. The study also showed a domain like structure with a hierarchical organization of vortices over pinning centers. Each domain is associated with different filling fractions. Such a multi-domain like vortex configuration described in the theoretical study [20] may also exhibit a heterogeneous pinning character. In one of our recent detailed studies, [29] detailed maps of the shielding current distribution inside the patterned region upon reducing the applied field from a high field value suggest regions with different critical currents co-habiting within the patterned region, which supports the recent theoretical scenario of vortex phase separation [20] inside the patterned region. The above scenarios pave the way for future investigations into the peculiarities of the vortex configurations possible inside the nanopatterned regions and also ways to enhance pinning.

## 4. Conclusion

In conclusion in a BSCCO single crystal patterned with a periodically spaced blind hole array we have identified signatures of enhanced flux shielding and heterogeneous pinning response of the vortex configuration inside the patterned region at low $H$ below $B_\phi$. We have identified the regime where the pinning from the blind hole array dominates over the intrinsic pinning in the sample and shown that in this regime, large field gradients are sustained near the periphery of the patterned region while at the same time flux density near the center of the patterned region increases with increasing $H$. The distribution of local field gradients and shielding current density across the patterned region is also found to be non-uniform in the dilute vortex density regime of our measurements. We believe our work paves the way for future explorations into the nature of vortex configurations possible in nanopatterned systems.

**Acknowledgments**

SSB acknowledges FIB related staff and help from Dr. Amit Banerjee and funding support from JSPS, Japan- DST India and Indo – Spain joint research projects.

**References**

[1]   Blatter G, Feigel'man M V, Geshkenbein V B, Larkin A I and Vinokur V M 1994 *Rev. Mod. Phys.* **66** 1125
[2]   Giamarchi T and Le Doussal P 1994 *Phys. Rev. Lett.* **72** 1530; 1995 *Phys. Rev. B* **52** 1242
[3]   Fisher M P A 1989 *Phys. Rev. Lett.* **62** 1415
[4]   Córdoba R, Baturina T I, Sesé J, Mironov A Yu, De Teresa J M, Ibarra M R, Nasimov D A, Gutakovskii A K, Latyshev A V, Guillamón I, Suderow H, Vieira S, Baklanov M R, Palacios J J and Vinokur V M 2013 *Nature Communications* **4** 1437
[5]   Shaw Gorky, Bag Biplab, Banerjee S S, Suderow Hermann and Tamegai T 2012 *Supercond. Sci. Technol.* **25** 095016
[6]   Shaw Gorky, Sinha Jaivardhan, Mohan Shyam and Banerjee S S 2010 *Supercond. Sci. Technol.* **23** 075002
[7]   Shaw Gorky, Mandal Pabitra, Bag Biplab, Banerjee S S, Tamegai T and Suderow Hermann 2012 *Appl. Surf. Sci.* **258** 4199
[8]   Banerjee S S, Shaw Gorky, Sinha Jaivardhan, Mohan Shyam and Mandal Pabitra 2010 *Physica C* **470** S817
[9]   Hebard A F, Fiory A T and Somekh S 1977 *IEEE Trans. Magn.* **1** 589
[10]  Baert M, Metlushko V V, Jonckheere R, Moshchalkov V V and Bruynseraede Y 1995 *Phys. Rev. Lett.* **74** 3269
[11]  Moshchalkov V V, Baert M, Metlushko V V, Rosseel E, van Bael M J, Temst K, Jonckheere R and Bruynserade Y 1996 *Phys. Rev. B* **54** 7385
[12]  Moshchalkov V V, Baert M, Metlushko V V, Rosseel E, van Bael M J, Temst K, Bruynserade Y and Jonckheere R 1998 *Phys. Rev. B* **57** 3615
[13]  Thakur A D, Ooi Shuuichi, Chockalingam Subbaiah P, Jesudasan John, Raychaudhuri Pratap and Hirata Kazuto 2009 *Appl. Phys. Lett.* **94** 262501
[14]  Tamegai T, Tominaga T, Naito D, Nakajima Y, Ooi S, Mochiku T and Hirata K 2010 *Physica C* **470** S784
[15]  Silhanek A V, Gutierrez J, Kramer R B G, Ataklti G W, Van de Vondel J and Moshchalkov V V 2011 *Phys. Rev. B* **83** 024509

[16]  Dai H, Liu J and Lieber C M 1994 *Phys. Rev. Lett.* **72** 748
[17]  Fasano Y and Menghini M 2008 *Supercond. Sci. Technol.* **21** 023001
[18]  Bean C P 1964 *Rev. Mod. Phys.* **36** 31
[19]  Cooley L D and Grishin A M 1995 *Phys. Rev. Lett.* **74** 2788
[20]  Laroshenko O, Rybalko V, Vinokur V M and Berlyand L 2013 *Scientific Report (Nature)* **3** 1758
[21]  Shaw Gorky, Mandal Pabitra, Banerjee S S and Tamegai T 2012 *New J. Phys.* **14** 083042
[22]  Mandal Pabitra, Chowdhury Debanjan, Banerjee S S and Tamegai T 2012 *Review of Sci. Instr.* **83** 123906
[23]  Soibel A, Zeldov E, Rappaport M, Myasoedov Y, Tamegai T, Ooi S, Konczykowski M and Geshkenbeink V B 2000 *Nature (London)* **406** 282
[24]  Banerjee S S, Soibel A, Myasoedov Y, Rappaport M, Zeldov E, Menghini M, Fasano Y, de la Cruz F, van der Beek C J, Konczykowski M and Tamegai T 2003 *Phys. Rev. Lett.* **90** 087004
[25]  Wijngaarden R J, Heeck K, Spoelder H J W, Surdeanu R and Griessen R 1998 *Physica C* **295** 177
[26]  Banerjee S S, Goldberg S, Soibel A, Myasoedov Y, Rappaport M, Zeldov E, de la Cruz F, van der Beek C J, Konczykowski M, Tamegai T and Vinokur V M 2004 *Phys. Rev. Lett.* **93** 097002
[27]  Bak P 1982 *Rep. Prog. Phys.* **45** 587
[28]  Brandt E H and Indenbom M 1993 *Phys. Rev. B* **48** 12893
[29]  Shaw Gorky, Banerjee S S, Tamegai T and Suderow Hermann 2016 *Supercond. Sci. Technol.* **29** 065021